\documentclass[iicol]{sn-jnl}

\usepackage{graphicx}%
\usepackage{multirow}%
\usepackage{amsmath,amssymb,amsfonts}%
\usepackage{amsthm}%
\usepackage{mathrsfs}%
\usepackage[title]{appendix}%
\usepackage{xcolor}%
\usepackage{textcomp}%
\usepackage{manyfoot}%
\usepackage{booktabs}%
\usepackage{algorithm}%
\usepackage{algorithmicx}%
\usepackage{algpseudocode}%
\usepackage{listings}%
\usepackage{svg}
\usepackage{float}
\theoremstyle{thmstyleone}%

\usepackage{dblfloatfix}

\theoremstyle{thmstyletwo}%

\theoremstyle{thmstylethree}%

\begin{document}

\title[Phase dynamics and dissipation in tunnel magnetic Josephson junctions]{Phase dynamics and dissipation in tunnel ferromagnetic Josephson junctions}

\author[1]{\fnm{F.} \sur{Calloni}}
\author[1]{\fnm{R.} \sur{Satariano}}
\author[2]{\fnm{R.} \sur{Ferraiuolo}}
\author[1]{\fnm{H.G.} \sur{Ahmad}}
\author[1]{\fnm{D.} \sur{Gatta}}
\author[1]{\fnm{E.} \sur{Raja}}
\author[1]{\fnm{G.} \sur{Santo}}
\author[1]{\fnm{G.} \sur{Serpico}}
\author[1]{\fnm{R.} \sur{Vidyasagar}}
\author[1]{\fnm{D.} \sur{Montemurro}}
\author[1,3]{\fnm{N.} \sur{Poccia}}
\author[4]{\fnm{A.} \sur{Vettoliere}}
\author[1]{\fnm{G.} \sur{Ausanio}}
\author[4]{\fnm{C.} \sur{Granata}}
\author[1]{\fnm{L.} \sur{Parlato}}
\author[1]{\fnm{G.P.} \sur{Pepe}}
\author[2]{\fnm{A.} \sur{Bruno}}
\author[1]{\fnm{F.} \sur{Tafuri}}
\author[5]{\fnm{D.} \sur{Massarotti}}

\affil[1]{University of Naples Federico II, Dipartimento di Fisica "Ettore Pancini", Via Cinthia, Naples, 80126, Italy}
\affil[2]{QuantWare, Elektronicaweg 10, Delft, 2628 XG, Netherlands}
\affil[3]{Leibniz Institute for Solid State and Materials Research, Dresden, 01069, Germany}
\affil[4]{National Research Council, Institute of Applied Sciences and Intelligent Systems (ISASI), Via Campi Flegrei 34, Pozzuoli, 80078, Italy}
\affil[5]{University of Naples Federico II, Dipartimento di Ingegneria Elettrica e delle Tecnologie dell'Informazione, Via Claudio 21, Naples, 80125, Italy}


\abstract{We investigate tunnel ferromagnetic Josephson junctions based on Superconductor-Insulator-thin superconductor-Ferromagnet-Superconductor multilayers. A comparative study of their electrodynamic properties is performed for junctions with niobium and aluminum (Al) electrodes, featuring different ferromagnetic interlayer materials 
and lateral dimensions ranging from the micrometric to the submicrometric scale. The parameters extracted from the fitting of the current-voltage characteristics using the tunnel junction microscopic model are found to be consistent with those independently estimated from switching current distribution measurements.
Submicrometric Al-based devices exhibit electrodynamic properties comparable to those implemented in state-of-the-art transmon qubits and display clear signatures of quantum phase diffusion. The strong agreement between transport modelling and escape dynamics establishes a robust framework for describing hybrid ferromagnetic Josephson junctions consistent with their energy scales and supports their potential integration into superconducting quantum and classical digital circuits.}





\maketitle

\section{Introduction}\label{sec1}
Magnetic Josephson junctions (MJJs) are key structures in the field of superconducting spintronics \cite{Buzdin_2005,Berget_2005,Eschrig_2011,Linder2015} and recently have attracted considerable attention as building blocks for both quantum and classical superconducting circuits \cite{Cai_2022,Birge2024,Feofanov_2010,Kim2024}. They have been proposed as energy-efficient cryogenic memories \cite{Larkin_2012,Goldobin_2013} and as passive $\pi$-shifters in superconducting digital and quantum circuits \cite{Birge2024,Feofanov_2010,Kim2024}. However, despite their potential, MJJs based on Superconductor-Ferromagnet-Superconductor (SFS) trilayers present several drawbacks that limit their integration into highly coherent circuits. They exhibit a very small critical current–normal resistance product $I_c R_N$, often only a few~$\mu$V or less \cite{Ryazanov_2001,Kontos2002,Weides2006}. In addition, these junctions are generally overdamped and affected by substantial quasiparticle dissipation \cite{Kato, Massarotti}, which degrades the performance of a wide variety of superconducting circuits \cite{Serniak2018, Connolly2024, Aquino2025}. For such applications, large $I_c R_N$ products and low damping are essential requirements \cite{Larkin_2012, Kato}. 

These parameters are crucial across different hybrid junction platforms. A notable example is found in cuprate-based junctions, where the observation of macroscopic quantum tunneling has paved the way for integrating these devices into quantum circuits \cite{Longobardi2005, Stornaiuolo2013, Confalone2025a, Confalone2025b}. \newline
Low-dissipation ferromagnetic junctions can be realized by inserting an additional insulating layer I between the superconducting electrode and the ferromagnetic barrier, forming SIFS Josephson junctions (JJs) \cite{Goldobin_2013,Weides2006APL,Weides2006PRL,Pfeiffer2008PRB,Bannykh2009PRB}, or also by inserting an additional superconductive interlayer thus forming SIsFS JJs \cite{Larkin_2012,Caruso_2018,Satariano_2024,vettoliere22,Satariano2025}. An alternative approach is based on the use of ferromagnetic insulator barriers, giving rise to SIfS junctions \cite{Senapati_2011,Massarotti2015,Ahmad2020}. 
Junctions incorporating insulating barriers help mitigate SFSs' limitations by restoring a more ideal tunnelling regime, while maintaining the magnetic control of the supercurrent \cite{Ahmad23_competition,Holmes-Hewett_2025}. Such tunnel hybrid structures have been also theoretically proposed for quantum devices, including quiet ferromagnetic flux qubits relying on anomalous $0$–$\pi$ transitions \cite{Kato}. More recently, tunnel ferromagnetic JJs have been proposed for the implementation of the so-called \emph{ferro-transmon} \cite{ahmad2022_ferrotrasmone,Massarotti2023Ferrotransmon}. In this configuration, the magnetic state of the ferromagnetic barrier acts as a nonvolatile control knob for the Josephson energy, enabling fast and local tuning by applying magnetic field pulses \cite{ahmad2022_ferrotrasmone,ahmad2025}.
 
In contrast to SIfS JJs, which face several limitations due to the limited availability of intrinsic ferromagnetic insulators, SIsFS JJs can be functionalized with a wide class of magnetic materials, such as PdFe and Ni$_{80}$Fe$_{20}$ (Permalloy: Py) \cite{Larkin_2012,Caruso_2018,Parlato}. At the same time, standard fabrication procedure of state-of-the-art superconducting circuits can be applied \cite{Larkin_2012,vettoliere22,Satariano2025}. Most importantly, the transport properties of SIsFS junctions can be engineered and optimized to meet the specific requirements of different applications \cite{Bakurskiy_2013}.  For instance, niobium-based devices enable fast switching for cryogenic memories compatible with Single Flux Quantum (SFQ) circuits  \cite{Larkin_2012,Caruso_2018,Parlato}, while aluminum-based devices are more naturally suited for quantum computing architectures \cite{vettoliere22,Satariano2025},  where aluminum technology remains the standard \cite{Siddiqi21}.
 
Understanding the phase dynamics of this class of hybrid junctions is thus essential for both fundamental physics and for their integration into superconducting circuits. Switching Current Distribution (SCD) measurements offer a powerful probe of the escape processes from the superconducting state, enabling the identification of thermal activation, macroscopic quantum tunnelling, and phase diffusion regimes \cite{Vion,Devoret1985,Martins1987,Kivioja2005}. In devices operating in low-to-moderate damping conditions, SCD measurements are particularly informative, revealing dissipation mechanisms quantified by the quality factors of the JJs \cite{PRL_Diagram}, which directly influence coherence in superconducting quantum devices. \newline
In this work, we investigate SIsFS junctions fabricated with both Nb \cite{Satariano_2024} and Al electrodes \cite{vettoliere22}, featuring different ferromagnetic barriers. For the aluminum samples, the F layer is permalloy, instead for the Nb-based micrometric junctions the ferromagnetic barrier is a (Ni$_{80}$Fe$_{20}$)$_{80}$Nb$_{20}$ alloy. For the Al-based devices, submicrometric junctions were realized to explore reduced-dimensionality effects relevant for scalable quantum technologies \cite{Satariano2025}. By performing a detailed comparative and systematic analysis of the phase dynamics, we correlate the electrodynamic parameters extracted from fitting the low-frequency current-voltage ($\text{I–V}$) characteristics, using the Tunnel Junction Microscopic (TJM) model, with those inferred from SCD measurements \cite{Massarotti2015,Ahmad2020,ahmad2024}. We can therefore probe the dissipation mechanisms across different materials and geometries, providing a comprehensive picture of the electrodynamic behavior of hybrid SIsFS junctions and assessing their suitability for future superconducting quantum and digital circuits.
\newpage

\section{Fabrications and Methods}\label{sec2}
Two distinct fabrication processes were employed depending on the lateral dimensions of the SIsFS junctions. 
For the micrometric SIsFS junctions, the Nb/AlO$_x$/Nb (or Al/AlO$_x$/Al) trilayer was patterned using optical lithography and a lift-off procedure. The junction areas were then defined by a selective anodization process, followed by insulation via SiO$_2$ deposition. Subsequently, soft Ar ion etching was employed to clean the Nb (or Al) surface, removing approximately 10 nm of the thin superconducting interlayer. This step precedes the ferromagnetic layer deposition, which utilizes a lift-off technique combined with magnetron sputtering. Finally, the S counter-electrode was deposited through a further d.c. sputtering and lift-off process, resulting in the complete SIsFS structure. Although the ferromagnetic barrier and top electrode are deposited in subsequent runs, the photolithographic masks and lift-off parameters were optimized to ensure precise spatial alignment, minimizing the occurrence of lateral geometry mismatches and associated fringe fields \cite{Nevirkovets2018,Nevirkovets2023}.
The Nb-based micrometric junctions (hereafter, referred to as Sample A) follow the multilayer sequence:
Nb (200\,nm) /AlO$_x$(2\,nm) /Nb(30\,nm) /(Ni$_{80}$Fe$_{20}$)$_{80}$Nb$_{20}$(6\,nm) /
Nb (350\,nm). For these junctions, the Py layer is doped with niobium to tailor its magnetic properties. The ferromagnetic layer was deposited by magnetron co-sputtering from a Py target and a Nb target, with applied powers of 120 W and 60 W, respectively. Magnetic characterizations performed on reference films confirmed that the magnetization of both standard Py and Nb-doped Py films show an in-plane magnetic anisotropy.
The presence of non-magnetic inclusions creates local discontinuities in the magnetic domains, acting as pinning sites for magnetic domain walls  \cite{QADER2017}. This mechanism promotes the magnetization process via domain wall motion and lowers the pinning energy, thereby reducing the external magnetic field required to fully magnetize the ferromagnetic layer \cite{Birge2024,Baek2014}. 
The Al-based micrometric junctions (Sample B) have the same nominal thickness layer as their Nb counterparts; however the ferromagnetic barrier is a 3 nm-thick Py layer: Al (200\,nm)/AlO$_x$(2\,nm)/Al (30\,nm)/Py (3\,nm)/Al (350\,nm). The fabrication process is specifically optimized for large-area JJs \cite{Parlato,Vettoliere22_nanomaterials}. \newline
Submicrometric Al-based junctions (Sample C)  were fabricated using the Manhattan process, employed for the realization of standard transmon qubits  \cite{Potts2001,Muthusubramanian_2024}.
In this method, the cross-shaped junction geometry is first defined by electron-beam lithography, followed by angled shadow evaporation of aluminum and \textit{in-situ} oxidation to form the Al/AlO$_x$/Al trilayer (SIs).  A second lithography step opens a 500\,nm~$\times$~500\,nm window above the junction area; after gentle ion milling, a 3\,nm-thick Py layer and the top Al electrode are deposited, completing the SIsFS stack. A final step defines the electrical contact for the wiring. This process yields submicrometric junctions with thicknesses Al (20\,nm)/AlO$_x$ (2\,nm)/Al (25\,nm)/Py (3\,nm)/Al (70\,nm) \cite{Satariano2025}. Owing to the fabrication process, the lateral dimensions of the FS electrode are smaller than those of the underlying SIs junction area in order to avoid electrical shorts between the bottom and top superconducting layers. As a consequence, fringe magnetic fields may arise at the edges of the ferromagnetic layer and could potentially influence the junction characteristics, as discussed in Refs. \cite{Nevirkovets2018,Nevirkovets2023}. 
\begin{figure*}[!]
    \centering
    \includegraphics[width=0.8 \textwidth]{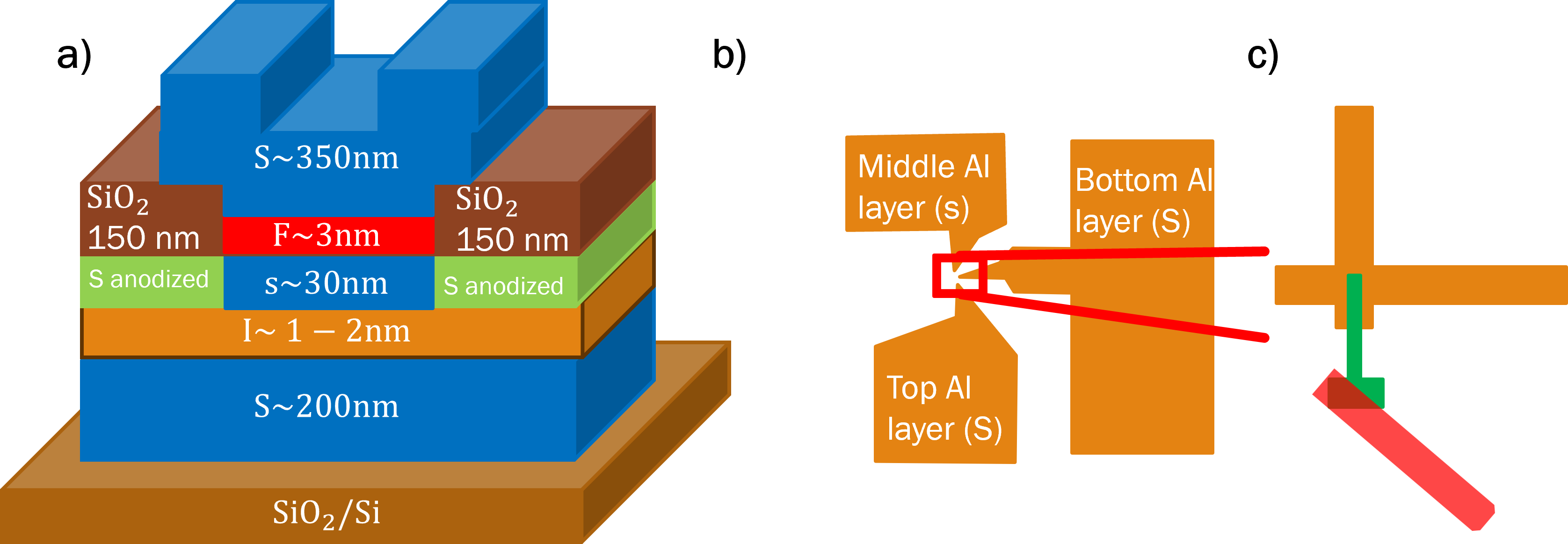}
    \caption{ \textbf{a)} Sketch of Superconductor/Ferromagnet/thin superconductor/Ferromagnet/Superconductor Josephson junctions (SIsFS JJs) with diameter ranging between 3 and 10$\mu$ m. \textbf{b)} Sketch of the design of the multilayer submicrometric SIsFS JJ. The pads (in orange) allow separate measurements of the SIs, sFS, and multilayer SIsFS junctions. \textbf{c)} Close-up of the junction area: the SIs part is shown in orange, the window for the FS deposition in green, and the aluminum connection between the junction and the pad in red.}
    \label{SEM}
\end{figure*}

The SIsFS junctions were measured by thermally anchoring the samples to the mixing chamber of a Triton 400 dry dilution refrigerator, equipped with customized low-noise filtering stages \cite{Salvoni2021}. For the acquisition of the I-V characteristics, the junctions are current-biased using a low-frequency triangular waveform (approximately 1\,Hz), and the voltage response is recorded through a differential amplifier. For SCD measurements, a voltage threshold is set close to the zero voltage state in order to accurately detect the actual switching event and track its occurrence over time. For each temperature, a total of $5 \times 10^3$ events are acquired. 
The switching current counts $N_I$ allow to estimate the switching current probability density distribution $P(I)$.
The mean switching current value ($I_{\text{mean}} = \sum_i I_i \frac{N_{I_i}}{N_{\text{tot}}}$) and its variance ($\sigma^2= \sum_i (I_i - I_{\text{mean}})^2 \frac{N_{I_i}}{N_{\text{tot}}}$) are determined as the first and second moments of the distribution, where $N_{\text{tot}}$ is the total number of counts and $I_i$ are the switching current values. Notably, the second moment enables the estimation of the distribution width $\sigma$, providing key information about the stochastic nature of the switching process \cite{Vion,Devoret1985,Martins1987,Kivioja2005}.

\section{Results}\label{sec3}

\begin{figure}
\centering
\includegraphics[width=8cm]{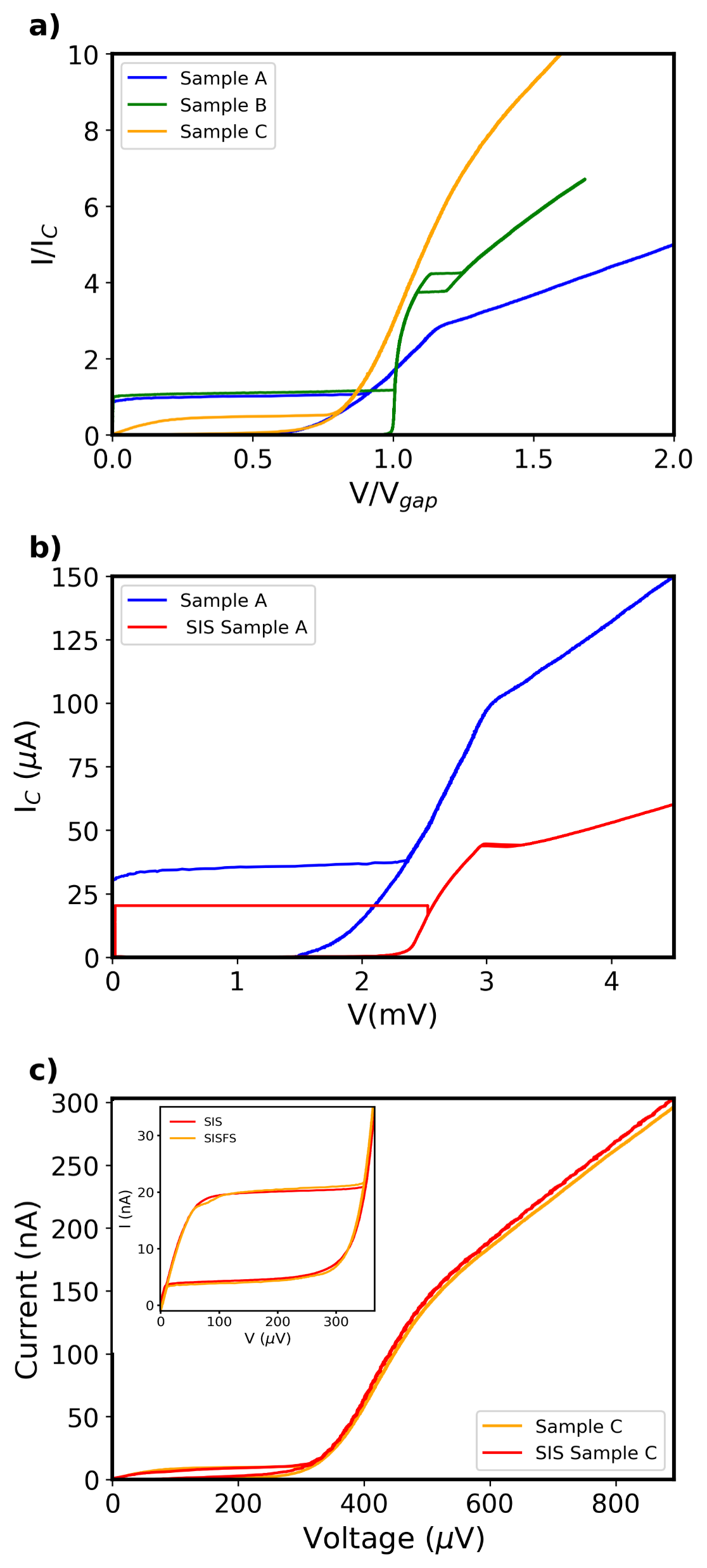}
\caption{Comparison of the Current-Voltage ($I$–$V$) characteristics at 10 mK. 
\textbf{a) }$I$–$V$ normalized with respect to the critical current $I_c$ and the superconducting gap voltage $V_{\mathrm{gap}}$, respectively. Sample A is shown in blue, sample B in green, and sample C in orange. \textbf{b) }Comparison between sample A (square junction with lateral size of 10 $\mu$m) and a reference niobium SIS junction (square junction with lateral size of 4 $\mu$m). \textbf{c)} Comparison between Sample C (square junction with a 500 nm lateral size) and aluminum SIS reference junction of comparable geometrical dimensions. The inset shows a magnification of the $I$–$V$ characteristics in the region close to the critical current $I_c$. \label{fig:LongRange}}
\end{figure} 
\begin{table*}[b]
\centering
\caption{Parameters of the samples measured at $T = 10$ mK. $A$ is the area of the junction, $J_c$ the critical current density, $I_cR_n$ the product of critical current and normal resistance, $V_{\mathrm{gap}}$ the voltage gap, $E_J$ the Josephson energy, $C_s$ the specific capacitance and $Q$ the geometric quality factor. Samples A, B and C correspond to micrometric Nb-based, micrometric Al-based and submicrometric Al-based SIsFS junctions, respectively.}
\label{Tabella1}
\begin{tabular}{c c c c c c c c }
\hline
Sample & $A$ ($\mu$m$^2$) & $J_c$ (A/cm$^2$) & $I_cR_n$ (mV) & $2V_{\mathrm{gap}}$ (mV)& $E_J$ ($\mu$eV) & $C_s$ (fF/$\mu$m$^2$) & $Q$ \\
\hline
A  & 100 & 30& 0.9& 2.5 & $60 \times 10^3$& 5 & 10\\
B & 13 & 0.5 & 0.08 & 0.45 & 130 & 4 & 12 \\
C & 0.25 & 8 & 0.07 & 0.44 & 40 & 4 & 40 \\
\hline
\end{tabular}
\end{table*}
In Figure \ref{fig:LongRange}a), we present the I-V characteristics for the samples A (blue curve), B (green curve), and C (orange curve). These samples differ in the composition of the ferromagnetic barrier and the superconducting electrodes, as detailed in the previous Section \ref{sec2}. The I-V curves are normalized with respect to the critical current $I_c$ and the gap voltage $V_{\mathrm{gap}}$, respectively.  For sample A, the critical current is on the order of tens of microamper ($I_c=32 \  \mu A$), whereas for the aluminum junctions the critical current is on the order of tens of  nanoampere ($I_c= 65$  nA for sample B and $I_{c}=20$ nA for sample C, respectively). The gap voltage $V_{\mathrm{gap}}$ was determined by numerically differentiating the $I$–$V$ curves and identifying the peak in $dI/dV$, yielding $2V_{\mathrm{gap}} \sim 440~\mu\mathrm{V}$ for aluminum and  $2V_{\mathrm{gap}} \sim 2.5~\mathrm{mV}$ for niobium.  
In figure \ref{fig:LongRange}b)  and c) the I-V characteristics are reported along with their non-magnetic SIS counterparts (red curves), in particular for sample A the SIS has a smaller area;  for sample B we refer to Ref. \cite{vettoliere22}.  The comparison with the I-V curves of the SIS JJs clearly highlights the tunneling behavior of these SIsFS junctions. 

The transport behavior of the SIsFS JJs can be discussed within the theoretical framework proposed by Bakurskiy \textit{et al.} \cite{Bakurskiy_2013}. According to this model, when the thickness of the intermediate superconducting layer $d_s$ exceeds the critical value $d_{sc}$, defined as the minimum thickness required for the superconductivity to persist in an sF bilayer, the pair potential $\Delta$ in the $s$ layer approaches the bulk value. Under these conditions, the SIsFS structure effectively behaves as a series combination of a tunnel SIs junction and a ferromagnetic sFS junction. In the limit of a high-resistance insulating barrier, the critical current of the tunnel SIs part $I_c^{\text{SIs}}$ is significantly smaller than that of the sFS side, $I_c^{\text{sFS}}$. Consequently, as shown in Figure \ref{fig:LongRange}, the $I$-$V$ characteristics are governed by the SIs component, exhibiting a clear tunnel-type behavior. \newline 
The experimental results reveal a more subtle scenario. For sample A, we observe a suppression of the critical current-normal resistance product $I_c R_N$  of  approximately 25$\%$ with respect to its SIS counterparts, whereas for samples B and C the  I-V curves coincide with the SIS ones \cite{vettoliere22,Satariano2025}. This small discrepancy can be attributed to the distinct properties of the Nb and Al interlayers. In sample A, the intermediate layer thickness ($d_s = 30$ nm) is larger than the superconducting coherence length in the dirty limit ($\xi_{Nb} \approx $ 10 nm). While the resulting ratio $d_s/\xi_{\text{Nb}} \approx 3$ ensures that the s layer is thick enough to establish a series regime, a reduction of the  $I_c R_N$  product is still observed compared to SIS reference values. This suppression suggests a weakening of the superconductivity in the s interlayer due to the proximity effect with the Py barrier, which induces a local reduction of the gap $\Delta_s$ relative to its bulk value and results in a higher dissipation in the subgap branch compared to the SIS I-V curve \cite{Bakurskiy_2013}. 

Regarding the samples B and C, they exhibit I-V curves that overlap with their SIS counterparts, thereby preserving all fundamental electrodynamic features. 
In aluminum thin-film, it is well known that the gap voltage $\Delta$ increases as the thickness decreases, due to structural and phonon-mediated effects \cite{Marchegiani_2022}. Consequently, the superconducting gap in these aluminum-based samples does not degrade, even in the presence of the Py layer. This point is further supported by the observation that although the s-interlayer in set C is thinner than in set B, the transport properties of the SIS component remain well preserved.
Moreover, the inverse proximity effect and the resulting spin diffusion across the S/F interface are strongly affected by the interface transparency \cite{Tsymbal_2007}. In Nb-based SIsFS junctions, previous investigations demonstrated a strong S/F coupling, evidenced by a lack of the magnetic hysteresis in the Fraunhofer patterns at very low temperatures \cite{Satariano_2024}. This strong coupling facilitates spin diffusion into the Nb s-layer, contributing to the observed reduction of the $I_c$$R_N$ product. Conversely, Al-based SIsFS junctions preserve a conventional magnetic hysteresis in their Fraunhofer modulation \cite{vettoliere22}, suggesting a much weaker coupling at the Al/Py interface. This weaker interaction can be attributed to the potential formation of a thin native AlO$_x$ layer at the s/F interface, which partially decouples the superconducting Al layer from the ferromagnetic barrier. A similar shielding mechanism due to interfacial oxides has been reported for Al films in contact with ferromagnetic insulators \cite{EuS_PRB, EuS_nature_communication}, effectively preventing the degradation of the superconducting gap via the proximity effect.


Despite these differences in material properties, all investigated junctions operate within the series regime predicted by the Bakurskiy model, confirming the versatility of the SIsFS architecture for both power-efficient digital and quantum electronics. The product $I_cR_N$ of the niobium junctions falls within the range suitable for cryogenic magnetic memories and SFQ circuits ($\approx$ 1 mV) \cite{Larkin_2012,Mukhanov2023}. On the other side, the aluminum junctions B and C exhibit lower $I_cR_N$ values, consistent with Josephson junctions $E_J $ employed in transmon qubits~\cite{Koch,Wisne_2024}, since for low values of $E_J (\sim 40\ \mu$eV) phase diffusion effects come into play, providing a reduction of the measurable critical current values \cite{Kivioja2005,Iansiti}.
Furthermore, introducing additional ferromagnetic layers and interfaces inevitably raises questions regarding the performance of these SIsFS JJs in microwave regime. In particular, non-ideal interfaces and amorphous oxide layers are well-known sources of two-level systems (TLSs), which can couple to the junction phase and introduce charge or flux noise \cite{Muller_2019, Paladino_2014}. While the presence of the F layer and the additional S/F interfaces could potentially increase the density of these fluctuators, the fact that our Al-based SIsFS junctions retain ideal SIS-like tunneling characteristics, without anomalous subgap leakage, suggests that the impact of additional TLS- and quasiparticle-induced dissipation is limited in d.c. regime. Nevertheless, integrating these SIsFS JJs into superconducting quantum circuits and characterizing their coherence properties will be essential in future optimization efforts to distinguish noise originating from the ferromagnetic barrier from other dominant noise sources in standard transmon architectures, such as the Purcell effect, quasiparticles, and flux noise \cite{Ahmad23_competition,ahmad2022_ferrotrasmone}. \newline
Table~\ref{Tabella1} summarizes the parameters estimated at $T=10$~mK for the analysed junctions. Here, as a first approximation, we have derived the junction specific capacitance $C_s$ following the empirical relation \cite{evaluation_C}: \(1/ C_{s} (\textrm{cm}^{2}/ \mu \textrm{F}) = 0.2 - 0.0043\log_{10} J_{c} (\textrm{kA}/\textrm{cm}^{2}) \), where $J_c$ is the measured critical current density.  From this value, a first estimate of the quality factor \(Q = \sqrt{\frac{2 e I_c R_N^2 C}{\hbar}}\) can be inferred, where $R_N$ is the normal resistance extracted from the experimental I-V characteristics. 
A more accurate determination requires adopting a two–quality–factor model for the junction phase dynamics.
In the tilted-washboard picture \cite{Barone}, the phase particle oscillates in a metastable minimum at the plasma frequency $\omega_p$,  in the superconducting state, whereas in the resistive state it evolves at $\omega \sim 0$ \cite{Kautz}. High-frequency dissipation at $\omega \sim \omega_p$, relevant during escape phenomena, is dominated by the electromagnetic environment, defining the high-frequency damping $Q_1$ \cite{Kautz,Martinis}. 
Low-frequency dissipation, associated with the subgap branch of the I–V curves, determines the low-frequency quality factor $Q_0$, governed by the subgap resistance $R_\mathrm{sg}$ \cite{Kautz,Martinis,Chen,Kirtley,Cristiano}. The TJM model provides a microscopic description of a JJ within the tunnelling Hamiltonian formalism \cite{Barone} and is widely used for SQUIDs and SFQ circuits \cite{Ahmad2020,Semenov,Odintsov, Polonsky}. The TJM is therefore an effective tool for extracting $Q_0$ and separating environmental from intrinsic dissipative contributions. \newline

\begin{figure}
\centering
\includegraphics[width=7.5cm]{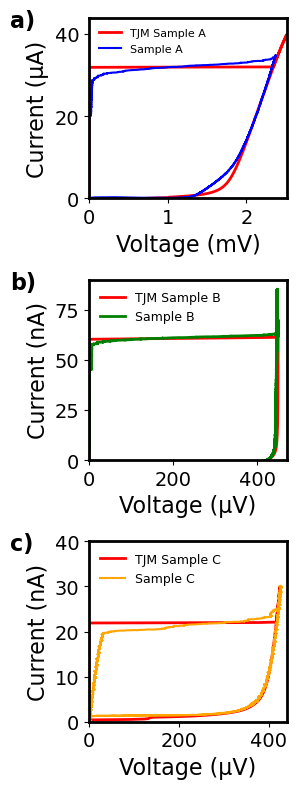}
\caption{\label{figTJM} TJM fit (red curve) and experimental I-V characteristics at 10 mK for \textbf{a)} sample A (blue curve) ; \textbf{b)}  sample B (green curve) and \textbf{c)} sample C (orange curve) .}
\end{figure}
TJM simulations were performed using the \texttt{PSCAN2} software \cite{sito}, which computes the time-averaged voltage $V(I)$ for a JJ embedded in a circuit. The parameters determining the I–V curve include $I_c$, the Stewart–McCumber parameter $\beta_C=Q_0^2$,  $V_g$, the ratio $I_c R_N/V_g$, and $R_N/R_\mathrm{sg}$. The measured $I_c$, $R_N$, and $V_g$ are fixed input quantities, while $\beta_C$ and $R_N/R_\mathrm{sg}$ serve as fitting parameters. More details on the fitting procedure can be found in Ref. \cite{Ahmad2020}. The accuracy of the fit is evident from its ability to reproduce the subgap branch of the I-V characteristics in Figure \ref{figTJM}.  From this estimate of $Q_0$, the corresponding junction capacitance and charging energy can be determined. For sample A, $\beta_C \sim 50$, corresponding to a  junction capacitance $C \sim 300~\mathrm{fF}$.  The quality factor $Q_0 $  is found to be 7, in contrast to the value of 10  for the standard Nb-based SIS junction \cite{Satariano2025}. This discrepancy can be attributed to the proximity effect affecting the s interlayer, which leads to metallic states in the subgap branch.  For sample B, $\beta_C \sim 100$, giving $Q_0 \sim 10$ and $C \sim 300~\mathrm{fF}$; finally for sample C , the parameter $\beta_C \sim 1$, which corresponds to a junction capacitance $C \sim 2~\mathrm{fF}$ and a charging energy $E_C \sim 40~\mu\mathrm{eV}$, as reported in Table~\ref{Capacità}. Despite the different materials, sample A and B exhibit comparable capacitances, as expected for tunnel barriers with similar specific capacitance.
\begin{table}[h!]
\centering
\renewcommand{\arraystretch}{1.4}
\caption{\label{Capacità} Quality factors $Q_0$ (from TJM fits) and $Q_1$ (from SCD measurements). The capacitance $C$ and the charging energy $E_C$ at $T = 10$ mK are determined from the $Q_0$ values.}
\begin{tabular}{|c|c|c|c|c|}
\hline
\textbf{JJs} & \textbf{Q$_0$} & \textbf{Q$_1$} & \textbf{C (fF)} & \textbf{E$_C$ ($\mu$eV)} \\
\hline
A & 7 & 1.5 & 250 & 0.3 \\
B & 10 & 1.3 & 300 & 0.3 \\
C & 1 & $<1.0$& 2 & 40 \\
\hline
\end{tabular}
\end{table}

\begin{figure*}[!] 
\centering
\includegraphics[width=1\textwidth]{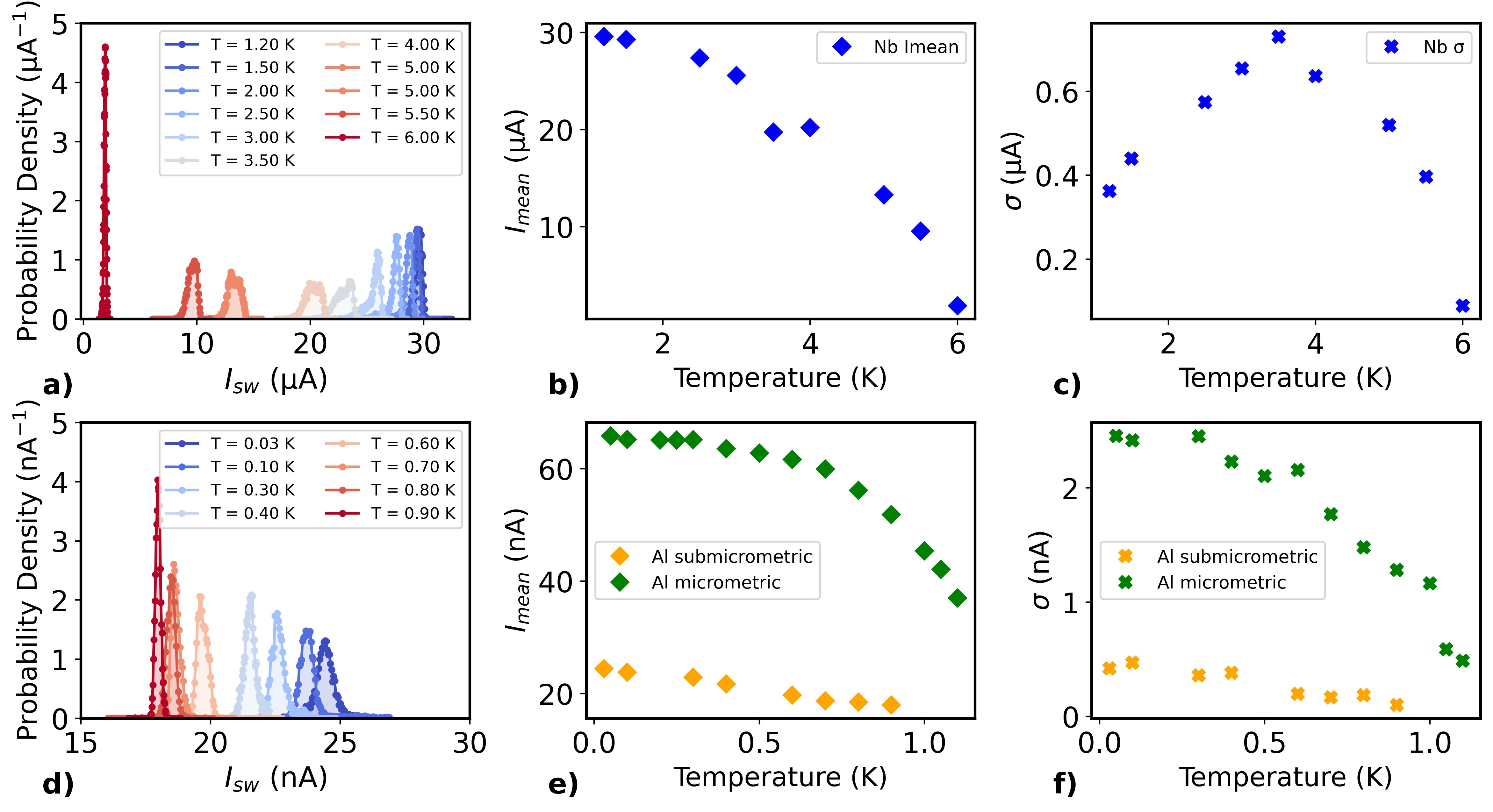}
\caption{\textbf{a)} Switching Current Distributions (SCDs) as a function of temperature for Sample A (10 mK--6 K). \textbf{b)} Temperature dependence of $I_{\mathrm{mean}}$ and \textbf{c)} $\sigma$ for Sample A. \textbf{d)} SCDs as a function of temperature for Sample C (10 mK-0.9 K). \textbf{e)} Temperature dependence of $I_{\mathrm{mean}}$ and \textbf{f)} $\sigma$ for Sample B (orange) and Sample C (green).}
\label{ScdNb}
\end{figure*} 
A detailed analysis of the the escape processes of the junctions has been carried out by performing SCD measurements. The SCDs as a function of temperature for Sample A are presented in Figure ~\ref{ScdNb}a), from which the temperature dependence of the mean switching current $I_{sw}$ and the standard deviation $\sigma$ have been derived in Figure ~\ref{ScdNb}b) and c). For this sample, $\sigma$ initially increases with temperature and subsequently decreases above a characteristic temperature $T^* \approx 3.5~\mathrm{K}$, indicating a crossover from thermal activation to a regime dominated by multiple retrapping and escape processes \cite{Kivioja2005}. At this temperature, the ratio $k_B T/ E_J$ is approximately $0.007$. From the phase–dynamics diagram reported in Ref.~\cite{PRL_Diagram}, we extract a value of $Q_1 \approx 1.5$. Considering that $I_c = 22~\mu\mathrm{A}$ at $3.5$ K and the presence of an on-chip resistance in series of $10 \ \Omega$,  we obtained a capacitance of $C \approx 250~\mathrm{fF}$, in close agreement with the value reported in Table \ref{Capacità}.  
For Sample B, the SCDs were previously reported in Ref.~\cite{ahmad2024}. 
For sample C, instead, the SCDs as a function of temperature are presented in Figure ~\ref{ScdNb}d).  Figs.~\ref{ScdNb}e) and f) show the temperature dependence of $I_{sw}$ and $\sigma$ for Samples B and C. Both aluminum samples exhibit the same temperature dependence, showing a $\sigma$ on the order of nanoamperes at the lowest temperatures and a monotonically decreasing trend as $T$ is increasing. Since sample A and sample B have similar capacitances, the  different behavior of $\sigma(T)$ primarily reflects their different Josephson energies $E_J$. In sample B, the reduced critical current $I_c = 65~\mathrm{nA}$ yields a significantly smaller $E_J$, such that the ratio $k_B T / E_J$ already reaches the value $\sim 0.006$ at $T = 10~\mathrm{mK}$, as a consequence, sample B operates in the phase diffusion regime  ($Q_1 \approx1.3$ \cite{ahmad2024}).  
At $T = 10~\mathrm{mK}$ sample C has $I_c = 20~\mathrm{nA}$, $\sigma  \sim 0.5~\mathrm{nA}$, $Q_1 < 1$, whose quality factor is also consistent with the phase diffusion temperature trend.
However, comparing Figure \ref{figTJM}b) and c) we do not observe for the sample B any resistance on the superconductive branch contrary to sample C.  
This provides a first signature that the escape process has a different origin, arising from the distinct $E_J/E_C$ ratios. In fact, sample A and B present $ E_J/E_C \gg 1$, whereas for sample C, $E_J/E_C \sim 1$.

\begin{figure}[h]
\centering
\includegraphics[width = 0.85\linewidth]{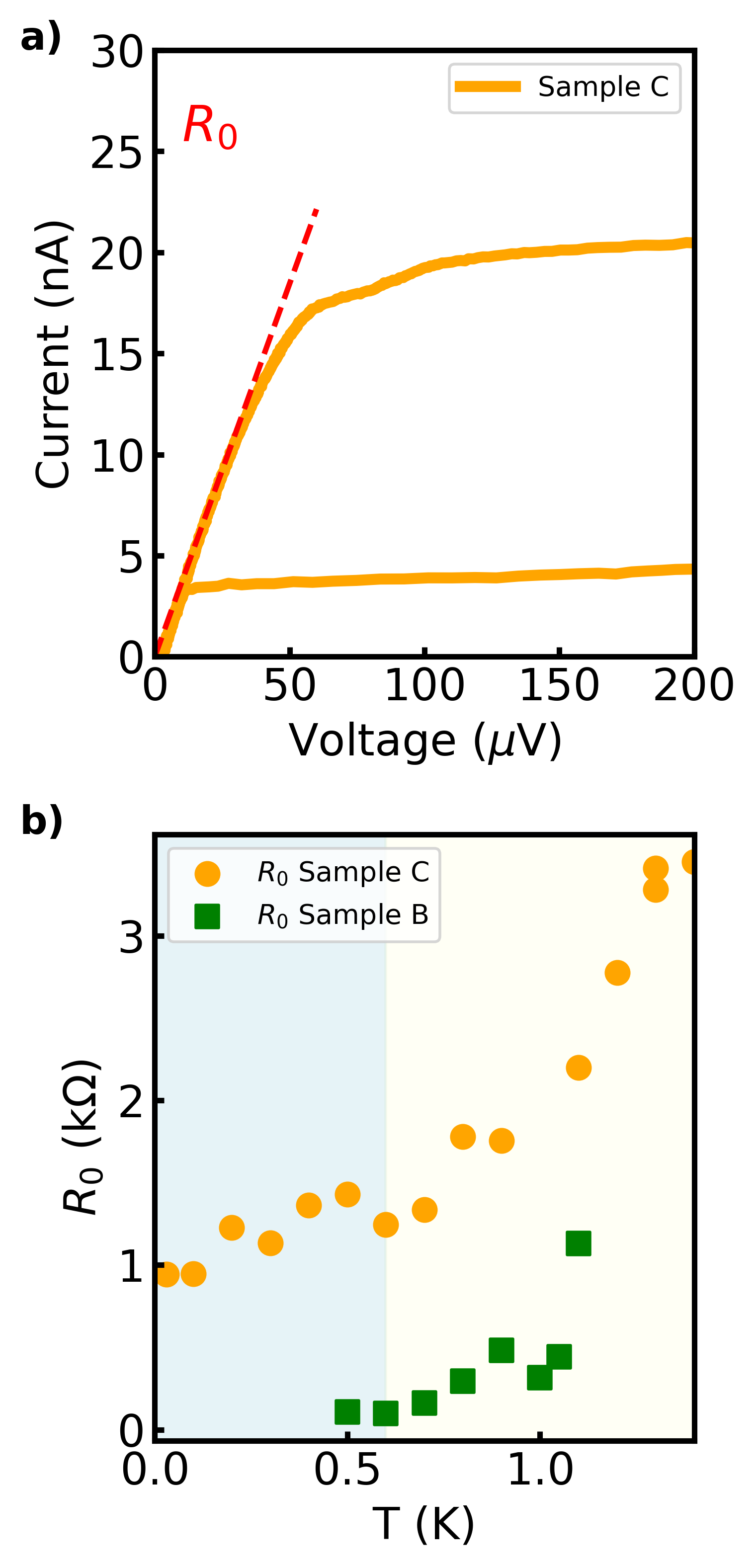}
\caption{\label{R0}\textbf{a)}: Zoom of the I-V characteristic of Sample C close to the zero voltage state, the red dashed line indicates the finite resistance $R_0$ of the low-voltage branch. \textbf{b)} Finite slope of the low-voltage branch of the I-V curves $R_0$ as a function of the temperature \textit{T} for the sample B (green square) and for the sample C (orange dots), where $R_0$ has been estimated through the linear fitting of the I-V on the low-voltage branch in Figure 4a. The light blue region indicates the quantum phase diffusion regime, while the yellow one indicates the phase diffusion regime activated by thermal fluctuations.}
\end{figure}

To point out this feature, we extract $R_0$ as a function of temperature by performing a linear fit of 
the low-voltage resistive branch close to the zero-voltage state, as reported in Figure \ref{R0}a). Due to the phase diffusion regime, a rounding of the I-V curves is physically expected \cite{Stornaiuolo2013}; therefore, the linear fit was strictly restricted to the low current bias region close to the zero current.
For the sample B, the I-V characteristics as a function of temperature, reported in Ref. \cite{ahmad2024},  do not show a clear resistance on the low-voltage branch till $T=0.6$ K. Due to the ratio $ E_J/E_C \gg 1$, the phase is strongly localized, but at $T=0.6$ K, $E_J \lesssim k_B T$ and thermal fluctuations strongly enhance the phase diffusion mechanism,  leading to the appearance and increase of $R_0$ with $T$ according to the Arrhenius law. For the sample C, the non-zero resistance at low temperatures originates from phase delocalization, due to $E_C/E_J \sim 1$, i.e., a Quantum Phase Diffusion (QPD) regime \cite{Stornaiuolo2013,Satariano2025,Vion,Iansiti}. 
Therefore, at low temperatures ($T \lesssim 0.6\ \mathrm{K}$),  $E_J >k_BT$, so the $R_0$ saturates  due to freezing out of the thermal fluctuations.
Further compelling evidence of the quantum nature of $R_0$ comes from the I-V characteristics as function of the magnetic field \cite{Satariano2025}. When a magnetic field is applied to suppress the Josephson energy $E_J$ at base temperature (where thermal fluctuations are negligible), a systematic and reproducible increase of the zero-bias resistance $R_0$ is observed. This strong dependence of $R_0$ on the magnetic tuning of $E_J$ demonstrates that the observed low-temperature resistance is a fundamental electrodynamic feature of the junction dynamics, rather than a spurious artifact of the fabrication process or external electromagnetic noise \cite{Satariano2025}.
As the temperature increases, thermal energy becomes comparable to the Josephson energy, so thermal fluctuations strongly enhance the phase diffusion mechanism, resulting in an exponential increase of $R_0$ with $T$. This temperature-dependent behavior provides a direct probe of the junction dynamics, clearly mapping the crossover from a low-dissipation regime (quantum/saturated $R_0$) to a thermally activated escape regime (increasing $R_0$) \cite{Iansiti}.
\newpage

\section{Outlook and Conclusions}\label{4}

The results of this study provide a comprehensive understanding of the transport properties and electrodynamics of SIsFS Josephson junctions. The TJM fit of the I-V characteristics demonstrates the model’s capability to accurately describe the subgap branch and low-frequency dynamics of the junctions. Simultaneously, SCD measurements provide detailed insights into phase dynamics, distinguishing between thermal activation and phase diffusion regimes.  This combined approach enables a more reliable extraction of key parameters, such as the junction capacitance, the charging energy $E_C$, the Josephson energy $E_J$, and the ratio $E_C/E_J$ , which are critical for superconducting circuit design \cite{Krantz_2019}. 
In Nb-based sample, the suppression of the $I_c R_N$ product underscores the impact of the proximity effect due to the ferromagnetic layer (Fig. \ref{fig:LongRange} b)). Nevertheless, these devices remain within the tunnel junction regime and preserve the intrinsic SFSs' memory properties, enabling fast switching suitable for cryogenic memories compatible with SFQ circuits \cite{Larkin_2012,Caruso_2018,Parlato}. Conversely, Al-based junctions (Samples B and C) preserve all the features of their SIS counterparts even at submicrometric scales (Fig. \ref{fig:LongRange} a) and c)).   
The results for submicron SIsFS junctions are comparable to those of standard Al-based junctions employed in conventional transmons, a finding further supported by the observation that they operate within the quantum phase diffusion limit \cite{Wisne_2024}.
 While direct coherence measurements are still required to fully assess their performance in quantum circuits, the observed low dissipation and clear tunneling behavior of the submicrometric Al-based SIsFS suggest their potential for superconducting quantum applications \cite{ Satariano2025,Ahmad23_competition}.
Ongoing work focuses on replacing the current ferromagnetic barrier with a softer ferromagnetic layer (e.g., Nb-doped permalloy), as the ferromagnetic material can be integrated \textit{ex-situ} without significantly affecting the junction electrodynamics. These findings support the potential for hybridizing aluminum-based quantum technologies with ferromagnetic elements in next-generation superconducting circuits.

\newpage

\section*{Acknowledgements}
This work has been supported by the Pathfinder EIC 2023 project "FERROMON-Ferrotransmons and Ferrogatemons for Scalable Superconducting Quantum Computers", the PNRR MUR project PE0000023-NQSTI, the PNRR MUR project CN-00000013-ICSC. R. S. acknowledges the project “MoQua–
Novel Magnetic Orders for superconducting Quantum Architectures”–Ricerca di Ateneo Linea C. 
\section*{Data Availability Statement}
The data that support the findings of this study are available from the corresponding author upon reasonable request.
\section*{Conflict of Interest statement}
The authors have no conflicts to disclose. 
\section*{Author contributions }
The original manuscript was written by F.C., R.S. and D.Ma. with contributions from all authors. F.C., R.S, H.G.A., D.G., E.R., G.Sa. and G.Se. carried out the measurements; F.C., R.S., D.G. and E.R. worked on the data analysis; R.F., D.Mo., A.V., G.A., C.G., L.P. and A.B. designed and realized the junctions; G.P. P., A.B, F.T. and D.Ma. contributed to the fundings. All authors discussed the results and commented on the manuscript.
  
\newpage
\bibliography{sn-bibliography}

@article{Nevirkovets2023,
    author = {Nevirkovets, I. P. and Mukhanov, O. A.},
    title = {Electrically controlled hybrid superconductor–ferromagnet cell for high density cryogenic memory},
    journal = {Applied Physics Letters},
    volume = {123},
    number = {7},
    pages = {072601},
    year = {2023},
    month = {08},
    issn = {0003-6951},
    doi = {10.1063/5.0165128},
    url = {https://doi.org/10.1063/5.0165128},
    eprint = {https://pubs.aip.org/aip/apl/article-pdf/doi/10.1063/5.0165128/18084515/072601_1_5.0165128.pdf},
}

@article{Muller_2019,
  title={Towards understanding two-level systems in amorphous solids: insights from quantum circuits},
  author={M{\"u}ller, Clemens and Cole, Jared H and Lisenfeld, J{\"u}rgen},
  journal={Reports on Progress in Physics},
  volume={82},
  number={12},
  pages={124501},
  year={2019},
  publisher={IOP Publishing}
}

@article{Paladino_2014,
  title={1/f noise: Implications for solid-state quantum information},
  author={Paladino, Elisabetta and Galperin, YM and Falci, Giuseppe and Altshuler, BL},
  journal={Reviews of Modern Physics},
  volume={86},
  number={2},
  pages={361},
  year={2014},
  publisher={APS}
}

@article{Tsymbal_2007,
  title = {Interface effects in spin-polarized tunneling and Josephson junctions},
  author = {Tsymbal, E. Y. and Mryasov, O. N. and LeClair, P. R.},
  journal = {Prog. Mater. Sci.},
  volume = {52},
  number = {2-3},
  pages = {401--420},
  year = {2007},
  doi = {10.1016/j.pmatsci.2006.10.009}
}

@article{Berget_2005,
  title = {Odd triplet superconductivity and related phenomena in superconductor-ferromagnet structures},
  author = {Bergeret, F. S. and Volkov, A. F. and Efetov, K. B.},
  journal = {Rev. Mod. Phys.},
  volume = {77},
  issue = {4},
  pages = {1321--1373},
  numpages = {0},
  year = {2005},
  month = {Nov},
  publisher = {American Physical Society},
  doi = {10.1103/RevModPhys.77.1321},
  url = {https://link.aps.org/doi/10.1103/RevModPhys.77.1321}
}

@article{Holmes-Hewett_2025,
doi = {10.1088/1361-6463/adf1b2},
url = {https://doi.org/10.1088/1361-6463/adf1b2},
year = {2025},
month = {aug},
publisher = {IOP Publishing},
volume = {58},
number = {34},
pages = {343001},
author = {Holmes-Hewett, W F and Miller, J D and Ahmad, H G and Granville, S and Ruck, B J},
title = {Rare-earth nitrides: fundamental advances and applications in cryogenic electronics},
journal = {Journal of Physics D: Applied Physics}
}

@article{Eschrig_2011,
  title = {Spin-polarized supercurrents for spintronics.},
  author = {Matthias Eschrig},
  journal = {Phys. Today 65, 43} ,
  year = {(2011)}
}

@article{Kim2024,
  author    = {Kim, S. and Abdurakhimov, L. V. and Pham, D. and others},
  title     = {Superconducting flux qubit with ferromagnetic {J}osephson $\pi$-junction operating at zero magnetic field},
  journal   = {Communications Materials},
  volume    = {5},
  pages     = {216},
  year      = {2024},
  doi       = {10.1038/s43246-024-00659-1}
}

@article{Linder2015,
  author    = {Jacob Linder and Jason W. A. Robinson},
  title     = {Superconducting spintronics},
  journal   = {Nature Physics},
  year      = {2015},
  volume    = {11},
  number    = {4},
  pages     = {307--315},
  doi       = {10.1038/nphys3242},
  url       = {https://doi.org/10.1038/nphys3242},
  issn      = {1745-2481}
}

@article{Ryazanov_2001,
  title = {Coupling of Two Superconductors through a Ferromagnet: Evidence for a $\ensuremath{\pi}$ Junction},
  author = {Ryazanov, V. V. and Oboznov, V. A. and Rusanov, A. Yu. and Veretennikov, A. V. and Golubov, A. A. and Aarts, J.},
  journal = {Phys. Rev. Lett.},
  volume = {86},
  issue = {11},
  pages = {2427--2430},
  numpages = {0},
  year = {2001},
  month = {Mar},
  publisher = {American Physical Society},
  doi = {10.1103/PhysRevLett.86.2427},
  url = {https://link.aps.org/doi/10.1103/PhysRevLett.86.2427}
}

@article{Kontos2002,
  author  = {Kontos, T. and Aprili, M. and Lesueur, J. and Grison, X.},
  title   = {Superconducting Proximity Effect at the Ferromagnetic Metallic Interface},
  journal = {Phys. Rev. Lett.},
  volume  = {89},
  pages   = {137007},
  year    = {2002}
}

@article{Serniak2018,
  author  = {Serniak, K. and Hays, M. and de Lange, G. and Diamond, S. and Shankar, S. and Burkhart, L. D. and Kou, A. and Frunzio, L. and Devoret, M. H. and Catelani, G.},
  title   = {Hot Nonequilibrium Quasiparticles in Transmon Qubits},
  journal = {Phys. Rev. Lett.},
  volume  = {121},
  pages   = {157701},
  year    = {2018}
}

@article{Martins1987,
  title = {Experimental tests for the quantum behavior of a macroscopic degree of freedom: The phase difference across a {J}osephson junction},
  author = {Martinis, John M. and Devoret, Michel H. and Clarke, John},
  journal = {Phys. Rev. B},
  volume = {35},
  issue = {10},
  pages = {4682--4698},
  numpages = {0},
  year = {1987},
  month = {Apr},
  publisher = {American Physical Society},
  doi = {10.1103/PhysRevB.35.4682},
  url = {https://link.aps.org/doi/10.1103/PhysRevB.35.4682}
}

@article{Weides2006APL,
  author  = {Weides, M. and Kemmler, M. and Kohlstedt, H. and Waser, R. and et al.},
  title   = {0$-\pi$ {J}osephson tunnel junctions with ferromagnetic barrier},
  journal = {Applied Physics Letters},
  volume  = {89},
  pages   = {122511},
  year    = {2006},
  doi     = {10.1063/1.2356104}
}

@article{Weides2006PRL,
  author  = {Weides, M. and Kohlstedt, H. and Waser, R.},
  title   = {High-quality ferromagnetic {J}osephson junctions with 0--$\pi$ ground states},
  journal = {Phys. Rev. Lett.},
  volume  = {97},
  pages   = {247001},
  year    = {2006}
}

@article{Weides2006,
  title = {0--$\pi$ Josephson tunnel junctions with ferromagnetic barrier},
  author = {Weides, M. and Kemmler, M. and Kohlstedt, H. and Waser, R. and Elsner, D. and Laufenberg, M. and Goldobin, E. and Koelle, D. and Kleiner, R.},
  journal = {Applied Physics Letters},
  volume = {89},
  number = {12},
  pages = {122511},
  year = {2006},
  doi = {10.1063/1.2356104}
}

@article{Bannykh2009PRB,
  title = {Josephson tunnel junctions with a ferromagnetic interlayer: From 0 to $\pi$ transition},
  author = {Bannykh, A. A. and Pfeiffer, J. and Meckbach, J. M. and Gaber, T. and Mints, R. G. and Goldobin, E. and Koelle, D. and Kleiner, R.},
  journal = {Physical Review B},
  volume = {79},
  number = {5},
  pages = {054501},
  year = {2009},
  doi = {10.1103/PhysRevB.79.054501}
}

@article{Pfeiffer2008PRB,
  author  = {Pfeiffer, J. and Kemmler, M. and Weides, M. and Kohlstedt, H.},
  title   = {Static and dynamic properties of {SIFS} {J}osephson junctions},
  journal = {Phys. Rev. B},
  volume  = {77},
  pages   = {214506},
  year    = {2008}
}

@article{Potts2001,
  author    = {Potts, A. and Routley, P. R. and Parker, G. J. and Baumberg, J. J. and de Groot, P. A. J.},
  title     = {Novel fabrication methods for submicrometer {J}osephson junction qubits},
  journal   = {Journal of Materials Science: Materials in Electronics},
  volume    = {12},
  number    = {4},
  pages     = {289--293},
  year      = {2001},
  doi       = {10.1023/A:1011279908265},
  url       = {https://doi.org/10.1023/A:1011279908265},
  issn      = {1573-482X}
}

@ARTICLE{Salvoni2021,
  author={Salvoni, Daniela and Parlato, Loredana and Ejrnaes, Mikkel and Mattioli, Francesco and Gaggero, Alessandro and Martini, Francesco and Ausanio, Giovanni and Massarotti, Davide and Montemurro, Domenico and Ahmad, Halima Giovanna and di Palma, Luigi and Tafuri, Francesco and Cristiano, Roberto and Pepe, Giovanni Piero},
  journal={IEEE Instrumentation \& Measurement Magazine}, 
  title={Demonstration of Single Photon Detection in Amorphous Molybdenum Silicide / Aluminium Superconducting Nanostrip}, 
  year={2021},
  volume={24},
  number={5},
  pages={69-74},
  doi={10.1109/MIM.2021.9491006}}

@article{Muthusubramanian_2024,
   title={Wafer-scale uniformity of {D}olan-bridge and bridgeless {M}anhattan-style {J}osephson junctions for superconducting quantum processors},
   volume={9},
   ISSN={2058-9565},
   url={http://dx.doi.org/10.1088/2058-9565/ad199c},
   DOI={10.1088/2058-9565/ad199c},
   number={2},
   journal={Quantum Science and Technology},
   publisher={IOP Publishing},
   author={Muthusubramanian, Nandini and Finkel, Matvey and Duivestein, Pim and Zachariadis, Christos and van der Meer, Sean L M and Veen, Hendrik M and Beekman, Marc W and Stavenga, Thijs and Bruno, Alessandro and DiCarlo, Leonardo},
   year={2024},
   month=feb, 
   pages={025006} 
}

@article{Koch,
  title = {Charge-insensitive qubit design derived from the {C}ooper pair box},
  author = {Koch, Jens and Yu, Terri M. and Gambetta, Jay and Houck, A. A. and Schuster, D. I. and Majer, J. and Blais, Alexandre and Devoret, M. H. and Girvin, S. M. and Schoelkopf, R. J.},
  journal = {Phys. Rev. A},
  volume = {76},
  issue = {4},
  pages = {042319},
  numpages = {19},
  year = {2007},
  month = {Oct},
  publisher = {American Physical Society},
  doi = {10.1103/PhysRevA.76.042319},
  url = {https://link.aps.org/doi/10.1103/PhysRevA.76.042319}
}

@article{Larkin_2012,
  title     = {Ferromagnetic {J}osephson switching device with high characteristic voltage},
  volume    = {100},
  issn      = {1077-3118},
  url       = {http://dx.doi.org/10.1063/1.4723576},
  doi       = {10.1063/1.4723576},
  number    = {22},
  journal   = {Applied Physics Letters},
  publisher = {AIP Publishing},
  author    = {Larkin, Timofei I. and Bol'ginov, Vitaly V. and Stolyarov, Vasily S. and Ryazanov, Valery V. and Vernik, Igor V. and Tolpygo, Sergey K. and Mukhanov, Oleg A.},
  year      = {2012},
  month     = may
}

@article{Goldobin_2013,
  title     = {Memory cell based on a $\phi$ {J}osephson junction},
  volume    = {102},
  issn      = {1077-3118},
  url       = {http://dx.doi.org/10.1063/1.4811752},
  doi       = {10.1063/1.4811752},
  number    = {24},
  journal   = {Applied Physics Letters},
  publisher = {AIP Publishing},
  author    = {Goldobin, E. and Sickinger, H. and Weides, M. and Ruppelt, N. and Kohlstedt, H. and Kleiner, R. and Koelle, D.},
  year      = {2013},
  month     = jun
}

@article{Buzdin_2005,
   title={Proximity effects in superconductor-ferromagnet heterostructures},
   volume={77},
   ISSN={1539-0756},
   url={http://dx.doi.org/10.1103/RevModPhys.77.935},
   DOI={10.1103/revmodphys.77.935},
   number={3},
   journal={Reviews of Modern Physics},
   publisher={American Physical Society (APS)},
   author={Buzdin, A. I.},
   year={2005},
   month=sep, pages={935–976} }

@article{Kato,
  title = {Decoherence in a superconducting flux qubit with a $\ensuremath{\pi}$-junction},
  author = {Kato, T. and Golubov, A. A. and Nakamura, Y.},
  journal = {Phys. Rev. B},
  volume = {76},
  issue = {17},
  pages = {172502},
  numpages = {4},
  year = {2007},
  month = {Nov},
  publisher = {American Physical Society},
  doi = {10.1103/PhysRevB.76.172502},
  url = {https://link.aps.org/doi/10.1103/PhysRevB.76.172502}
}

@article{Massarotti,
  title = {Electrodynamics of {J}osephson junctions containing strong ferromagnets},
  author = {Massarotti, D. and Banerjee, N. and Caruso, R. and Rotoli, G. and Blamire, M. G. and Tafuri, F.},
  journal = {Phys. Rev. B},
  volume = {98},
  issue = {14},
  pages = {144516},
  numpages = {9},
  year = {2018},
  month = {Oct},
  publisher = {American Physical Society},
  doi = {10.1103/PhysRevB.98.144516},
  url = {https://link.aps.org/doi/10.1103/PhysRevB.98.144516}
}

@article{Feofanov_2010,
  author    = {A. K. Feofanov and V. A. Oboznov and V. V. Bol'ginov and J. Lisenfeld and S. Poletto and V. V. Ryazanov and A. N. Rossolenko and M. Khabipov and D. Balashov and A. B. Zorin and P. N. Dmitriev and V. P. Koshelets and A. V. Ustinov},
  title     = {Implementation of superconductor/ferromagnet/superconductor $\pi$-shifters in superconducting digital and quantum circuits},
  journal   = {Nature Physics},
  year      = {2010},
  volume    = {6},
  number    = {8},
  pages     = {593--597},
  doi       = {10.1038/nphys1700},
  url       = {https://doi.org/10.1038/nphys1700},
  issn      = {1745-2481},
}

@article{EuS_PRB,
  title = {Spin-filter effect of ferromagnetic europium sulfide tunnel barriers},
  author = {Hao, X. and Moodera, J. S. and Meservey, R.},
  journal = {Phys. Rev. B},
  volume = {42},
  issue = {13},
  pages = {8235--8243},
  numpages = {0},
  year = {1990},
  publisher = {American Physical Society},
  doi = {10.1103/PhysRevB.42.8235},
  url = {https://link.aps.org/doi/10.1103/PhysRevB.42.8235}
}

@article{EuS_nature_communication,
  title = {Spin regulation in composite spin-filter barrier devices},
  author = {Miao, G. X. and Chang, J. and Assaf, B. A. and Heiman, D. and  Moodera, J. S. },
  journal = {Nat. Commun.},
  volume = {5},
  issue = {1},
  pages = {3682},
  year = {2014}, 
  doi = {10.1038/ncomms4682}
}

@article{Nevirkovets2018,
  title = {Memory Cell for High-Density Arrays Based on a Multiterminal Superconducting-Ferromagnetic Device},
  author = {Nevirkovets, I. P. and Mukhanov, O. A.},
  journal = {Phys. Rev. Appl.},
  volume = {10},
  issue = {3},
  pages = {034013},
  numpages = {6},
  year = {2018},
  month = {Sep},
  publisher = {American Physical Society},
  doi = {10.1103/PhysRevApplied.10.034013},
  url = {https://link.aps.org/doi/10.1103/PhysRevApplied.10.034013}
}

@article{vettoliere22,
    author = {Vettoliere, A. and Satariano, R. and Ferraiuolo, R. and Di Palma, L. and Ahmad, H. G. and Ausanio, G. and Pepe, G. P. and Tafuri, F. and Montemurro, D. and Granata, C. and Parlato, L. and Massarotti, D.},
    title = {Aluminum-ferromagnetic {J}osephson tunnel junctions for high quality magnetic switching devices},
    journal = {Applied Physics Letters},
    volume = {120},
    number = {26},
    pages = {262601},
    year = {2022},
    month = {06},
    issn = {0003-6951},
    doi = {10.1063/5.0101686},
    url = {https://doi.org/10.1063/5.0101686},
    eprint = {https://pubs.aip.org/aip/apl/article-pdf/doi/10.1063/5.0101686/16450324/262601_1_online.pdf},
}

@article{Satariano2025,
    author = {Satariano, R. and Ferraiuolo, R. and Calloni, F. and Ahmad, H. G. and Gatta, D. and Tafuri, F. and Bruno, A. and Massarotti, D.},
    title = {Submicrometer tunnel ferromagnetic {J}osephson junctions with transmon energy scale},
    journal = {Applied Physics Letters},
    volume = {127},
    number = {25},
    pages = {252601},
    year = {2025},
    month = {12},
    issn = {0003-6951},
    doi = {10.1063/5.0303194},
    url = {https://doi.org/10.1063/5.0303194},
    eprint = {https://pubs.aip.org/aip/apl/article-pdf/doi/10.1063/5.0303194/20848839/252601_1_5.0303194.pdf},
}

@article{Vettoliere22_nanomaterials,
  author       = {Vettoliere, Antonio and Satariano, Roberta and Ferraiuolo, Raffaella and Di Palma, Luigi and Ahmad, Halima Giovanna and Ausanio, Giovanni and Pepe, Giovanni Piero and Tafuri, Francesco and Massarotti, Davide and Montemurro, Domenico and Granata, Carmine and Parlato, Loredana},
  title        = {High-Quality Ferromagnetic {J}osephson Junctions Based on Aluminum electrodes},
  journal      = {Nanomaterials},
  year         = {2022},
  volume       = {12},
  number       = {23},
  pages        = {4155},
  doi          = {10.3390/nano12234155},
  url          = {https://doi.org/10.3390/nano12234155}
}

@article{Massarotti2023Ferrotransmon,
  author       = {Massarotti, D. and Ahmad, H. G. and Satariano, R. and Ferraiuolo, R. and Di Palma, L. and Mastrovito, P. and Serpico, G. and Levochkina, A. and Caruso, R. and Miano, A. and Arzeo, M. and Ausanio, G. and Granata, C. and Lucignano, P. and Montemurro, D. and Parlato, L. and Vettoliere, A. and Fazio, R. and Mukhanov, O. and Pepe, G. P. and Tafuri, F.},
  title        = {A feasible path for the use of ferromagnetic {J}osephson junctions in quantum circuits: the ferro-transmon},
  journal      = {Low Temperature Physics},
  year         = {2023},
  volume       = {49},
  number       = {7},
  pages        = {794--802},
  doi          = {10.1063/10.0019690},
  url          = {https://doi.org/10.1063/10.0019690}
}

@ARTICLE{Ahmad23_competition,
  author={Ahmad, H. G. and Brosco, V. and Miano, A. and Di Palma, L. and Arzeo, M. and Satariano, R. and Ferraiuolo, R. and Lucignano, P. and Vettoliere, A. and Granata, C. and Parlato, L. and Ausanio, G. and Montemurro, D. and Pepe, G. P. and Fazio, R. and Tafuri, F. and Massarotti, D.},
  journal={IEEE Transactions on Applied Superconductivity}, 
  title={Competition of Quasiparticles and Magnetization Noise in Hybrid Ferromagnetic Transmon Qubits}, 
  year={2023},
  volume={33},
  number={5},
  pages={1-6},
  doi={10.1109/TASC.2023.3243197}}

@article{Birge2024,
    author = {Birge, Norman O. and Satchell, Nathan},
    title = {Ferromagnetic materials for {J}osephson $\pi$ junctions},
    journal = {APL Materials},
    volume = {12},
    number = {4},
    pages = {041105},
    year = {2024},
    month = {04},
    issn = {2166-532X},
    doi = {10.1063/5.0195229},
    url = {https://doi.org/10.1063/5.0195229},
    eprint ={https://pubs.aip.org/aip/apm/articlepdf/doi/10.1063/5.0195229/19866826/041105_1_5.0195229.pdf}
}

@article{Siddiqi21,
    author = {Siddiqi, Irfan},
    title = {Engineering high-coherence superconducting qubits},
    journal = {Nature Reviews Materials},
    volume = {6},
    number = {10},
    pages = {875-891},
    year = {2021},
    doi = {10.1038/s41578-021-00370-4},
    url = {https://doi.org/10.1038/s41578-021-00370-4}
    
}

@article{Satariano_2024,
   title={Nanoscale spin ordering and spin screening effects in tunnel ferromagnetic {J}osephson junctions.},
   volume={5},
   url={ https://doi.org/10.1038/s43246-024-00497-1},
   DOI={10.1038/s42005-021-00783-1},
   number={67},
   journal={Communications Materials},
   author={Satariano, R. and Volkov, A.F. and Ahmad, H.G. et al.},
   year={2024},
   }

@article{QADER2017,
title = {The magnetic, electrical and structural properties of copper-permalloy alloys},
journal = {Journal of Magnetism and Magnetic Materials},
volume = {442},
pages = {45-52},
year = {2017},
issn = {0304-8853},
doi = {https://doi.org/10.1016/j.jmmm.2017.06.081},
url = {https://www.sciencedirect.com/science/article/pii/S0304885317307035},
author = {Makram A. Qader and Alena Vishina and Lei Yu and Cougar Garcia and R.K. Singh and N.D. Rizzo and Mengchu Huang and Ralph Chamberlin and K.D. Belashchenko and Mark {van Schilfgaarde} and N. Newman}
}

@article{Barone,
  author    = { A. Barone and G. Paterno}, 
  title     = {Physics and Application of
 the {J}osephson Effect},
  publisher = {John Wiley \& Sons, Ltd},
  year      = {1982},
  isbn      = {9783527602780},
  doi       = {10.1002/352760278X.ch11},
  url       = {https://onlinelibrary.wiley.com/doi/abs/10.1002/352760278X.ch11}
 }

@inbook{Tafuri,
author = {Massarotti, D. and Tafuri, F.},
year = {2019},
month = {09},
pages = {455-512},
title = {Phase Dynamics and Macroscopic Quantum Tunneling},
isbn = {978-3-030-20724-3},
doi = {10.1007/978-3-030-20726-7_11}
}

@article{Martinis,
  title = {Classical phase diffusion in small hysteretic {J}osephson junctions},
  author = {Martinis, John M. and Kautz, R. L.},
  journal = {Phys. Rev. Lett.},
  volume = {63},
  issue = {14},
  pages = {1507--1510},
  numpages = {0},
  year = {1989},
  month = {Oct},
  publisher = {American Physical Society},
  doi = {10.1103/PhysRevLett.63.1507},
  url = {https://link.aps.org/doi/10.1103/PhysRevLett.63.1507}
}

@article{Kautz,
  title = {Noise-affected {I-V} curves in small hysteretic {J}osephson junctions},
  author = {Kautz, R. L. and Martinis, John M.},
  journal = {Phys. Rev. B},
  volume = {42},
  issue = {16},
  pages = {9903--9937},
  numpages = {0},
  year = {1990},
  month = {Dec},
  publisher = {American Physical Society},
  doi = {10.1103/PhysRevB.42.9903},
  url = {https://link.aps.org/doi/10.1103/PhysRevB.42.9903}
}

@article{Baek2014,
  title = {Hybrid superconducting-magnetic memory device using competing order parameters},
  author = {Baek, Burm and Rippard, William H. and Benz, Samuel P. and  Russek, Stephen E. and Dresselhaus, Paul D},
  journal = {Nature Communications},
  volume = {5},
  issue = {1},
  pages = {3888},
   year = {2014},
  doi = {10.1038/ncomms4888}
  
}

@article{Chen,
    author = {Chen, Y. C. and Fisher, Matthew P. A. and Leggett, A. J.},
    title = {The return of a hysteretic {J}osephson junction to the zero‐voltage state: {I‐V} characteristic and quantum retrapping},
    journal = {Journal of Applied Physics},
    volume = {64},
    number = {6},
    pages = {3119-3142},
    year = {1988},
    month = {09},
    issn = {0021-8979},
    doi = {10.1063/1.341527},
    url = {https://doi.org/10.1063/1.341527},
    eprint = {https://pubs.aip.org/aip/jap/article-pdf/64/6/3119/18619455/3119_1_online.pdf},
}

@article{Kirtley,
  title = {Measurement of the Intrinsic Subgap Dissipation in {J}osephson Junctions},
  author = {Kirtley, J. R. and Tesche, C. D. and Gallagher, W. J. and Kleinsasser, A. W. and Sandstrom, R. L. and Raider, S. I. and Fisher, M. P. A.},
  journal = {Phys. Rev. Lett.},
  volume = {61},
  issue = {20},
  pages = {2372--2375},
  numpages = {0},
  year = {1988},
  month = {Nov},
  publisher = {American Physical Society},
  doi = {10.1103/PhysRevLett.61.2372},
  url = {https://link.aps.org/doi/10.1103/PhysRevLett.61.2372}
}

@article{Cristiano,
    author = {Cristiano, R. and Frunzio, L. and Nappi, C. and Castellano, M. G. and Torrioli, G. and Cosmelli, C.},
    title = {The effective dissipation in ${Nb/AlO_x/Nb}$ {J}osephson tunnel junctions by return current measurements},
    journal = {Journal of Applied Physics},
    volume = {81},
    number = {11},
    pages = {7418-7426},
    year = {1997},
    month = {06},
    issn = {0021-8979},
    doi = {10.1063/1.365282},
    url = {https://doi.org/10.1063/1.365282},
    eprint = {https://pubs.aip.org/aip/jap/article-pdf/81/11/7418/18694471/7418_1_online.pdf},
}

@article{Semenov,
    author = {V. K. Semenov and A. A. Odintsov and A. B. Zorin},
    title = {SQUID’85—Superconducting Quantum Interference Devices and Their Applications},
    journal = {Walter de Gruyter \& Co.,
 Berlin},
    year ={1985} 
}

@ARTICLE{Odintsov,
  author={Odintsov, A. and Semenov, V. and Zorin, A.},
  journal={IEEE Transactions on Magnetics}, 
  title={Specific problems of numerical analysis of the {J}osephson junction circuits}, 
  year={1987},
  volume={23},
  number={2},
  pages={763-766},
  doi={10.1109/TMAG.1987.1064907}}

@article{Polonsky,
    author = {S. V. Polonsky and V. K. Semenov and P. N. Schevchenko},
    title = {PSCAN:Personal superconductor circuit analyser},
    journal = {Supercond. Sci. Technol. 4, 667},
    year = {1991}
}

@article{Devoret1985,
  title = {Measurements of Macroscopic Quantum Tunneling out of the Zero-Voltage State of a Current-Biased {J}osephson Junction},
  author = {Devoret, Michel H. and Martinis, John M. and Clarke, John},
  journal = {Phys. Rev. Lett.},
  volume = {55},
  issue = {18},
  pages = {1908--1911},
  numpages = {0},
  year = {1985},
  month = {Oct},
  publisher = {American Physical Society},
  doi = {10.1103/PhysRevLett.55.1908},
  url = {https://link.aps.org/doi/10.1103/PhysRevLett.55.1908}
}

@article{Longobardi2005,
  title = {Macroscopic Quantum Tunneling in $d$-Wave
  ${YBa}_{2}{Cu}_{3}{O}_{7-\delta}$
  {J}osephson Junctions},
  author = {Bauch, T. and Lombardi, F. and Tafuri, F. and Barone, A. and Rotoli, G. and Delsing, P. and Claeson, T.},
  journal = {Phys. Rev. Lett.},
  volume = {94},
  issue = {8},
  pages = {087003},
  numpages = {4},
  year = {2005},
  month = {Mar},
  publisher = {American Physical Society},
  doi = {10.1103/PhysRevLett.94.087003},
  url = {https://link.aps.org/doi/10.1103/PhysRevLett.94.087003}
}

@article{Kivioja2005,
  author = {Kivioja, J. M. and Nieminen, T. E. and Claudon, J. and Buisson, O. and Hekking, F. W. J. and Pekola, J. P.},
  title = {Observation of Transition from Escape Dynamics to Underdamped Phase Diffusion in a {J}osephson Junction},
  journal = {Phys. Rev. Lett.},
  volume = {94},
  pages = {247002},
  year = {2005},
  doi = {10.1103/PhysRevLett.94.247002}
}

@article{Massarotti2015,
  author  = {Massarotti, D. and Pal, A. and Rotoli, G. and Longobardi, L. and Blamire, M. G. and Tafuri, F.},
  title   = {Macroscopic quantum tunnelling in spin filter ferromagnetic {J}osephson junctions},
  journal = {Nature Communications},
  year    = {2015},
  month   = {jun},
  volume  = {6},
  number  = {1},
  pages   = {7376},
  doi     = {10.1038/ncomms8376},
  url     = {https://doi.org/10.1038/ncomms8376}
}

@article{sito,
    author = {PSCAN2 software (http://www.pscan2sim.org/)}
}

@article{Bakurskiy_2013,
   title={Theoretical model of superconducting spintronic {SI}s{FS} devices},
   volume={102},
   ISSN={1077-3118},
   url={http://dx.doi.org/10.1063/1.4805032},
   DOI={10.1063/1.4805032},
   number={19},
   journal={Applied Physics Letters},
   publisher={AIP Publishing},
   author={Bakurskiy, S. V. and Klenov, N. V. and Soloviev, I. I. and Bol’ginov, V. V. and Ryazanov, V. V. and Vernik, I. V. and Mukhanov, O. A. and Kupriyanov, M. Yu. and Golubov, A. A.},
   year={2013},
   month=may }

@article{Mukhanov2023,
  author = {O. A. Mukhanov and others},
  title = {High density fabrication process for single flux quantum circuits},
  journal = {Applied Physics Letters},
  volume = {122},
  number = {21},
  pages = {212601},
  year = {2023},
  doi = {10.1063/5.0152552}
}

@article{Vion,
  title = {Thermal Activation above a Dissipation Barrier: Switching of a Small {J}osephson Junction},
  author = {Vion, D. and G\"otz, M. and Joyez, P. and Esteve, D. and Devoret, M. H.},
  journal = {Phys. Rev. Lett.},
  volume = {77},
  issue = {16},
  pages = {3435--3438},
  numpages = {0},
  year = {1996},
  month = {Oct},
  publisher = {American Physical Society},
  doi = {10.1103/PhysRevLett.77.3435},
  url = {https://link.aps.org/doi/10.1103/PhysRevLett.77.3435}
}

@ARTICLE{ahmad2025,
  author={Ahmad, Halima Giovanna and Ferraiuolo, Raffaella and Serpico, Giuseppe and Satariano, Roberta and Levochkina, Anna and Vettoliere, Antonio and Granata, Carmine and Montemurro, Domenico and Esposito, Martina and Ausanio, Giovanni and Parlato, Loredana and Pepe, Giovanni Piero and Bruno, Alessandro and Tafuri, Francesco and Massarotti, Davide},
  journal={IEEE Transactions on Applied Superconductivity}, 
  title={Towards Novel Tunability Schemes for Hybrid Ferromagnetic Transmon Qubits}, 
  year={2025},
  volume={35},
  number={5},
  pages={1-7},
  doi={10.1109/TASC.2025.3535674}}

@article{Ahmad2020,
  title = {Electrodynamics of Highly Spin-Polarized Tunnel {J}osephson Junctions},
  author = {Ahmad, H.G. and Caruso, R. and Pal, A. and Rotoli, G. and Pepe, G.P. and Blamire, M.G. and Tafuri, F. and Massarotti, D.},
  journal = {Phys. Rev. Appl.},
  volume = {13},
  issue = {1},
  pages = {014017},
  numpages = {10},
  year = {2020},
  month = {Jan},
  publisher = {American Physical Society},
  doi = {10.1103/PhysRevApplied.13.014017},
  url = {https://link.aps.org/doi/10.1103/PhysRevApplied.13.014017}
}

@article{Caruso_2018,
   title={Properties of Ferromagnetic {J}osephson Junctions for Memory Applications},
   volume={28},
   ISSN={1558-2515},
   url={http://dx.doi.org/10.1109/TASC.2018.2836979},
   DOI={10.1109/tasc.2018.2836979},
   number={7},
   journal={IEEE Transactions on Applied Superconductivity},
   publisher={Institute of Electrical and Electronics Engineers (IEEE)},
   author={Caruso, Roberta and Massarotti, Davide and Miano, Alessandro and Bolginov, Vitaliy V. and Hamida, Aymen Ben and Karelina, Liubov N. and Campagnano, Gabriele and Vernik, Igor V. and Tafuri, Francesco and Ryazanov, Valery V. and Mukhanov, Oleg A. and Pepe, Giovanni P.},
   year={2018},
   month=oct, pages={1–6} }

@article{Parlato,
    author = {Parlato, Loredana and Caruso, Roberta and Vettoliere, Antonio and Satariano, Roberta and Ahmad, Halima Giovanna and Miano, Alessandro and Montemurro, Domenico and Salvoni, Daniela and Ausanio, Giovanni and Tafuri, Francesco and Pepe, Giovanni Piero and Massarotti, Davide and Granata, Carmine},
    title = {Characterization of scalable {J}osephson memory element containing a strong ferromagnet},
    journal = {Journal of Applied Physics},
    volume = {127},
    number = {19},
    pages = {193901},
    year = {2020},
    month = {05},
    issn = {0021-8979},
    doi = {10.1063/5.0004554},
    url = {https://doi.org/10.1063/5.0004554},
    eprint = {https://pubs.aip.org/aip/jap/article-pdf/doi/10.1063/5.0004554/15244693/193901_1_online.pdf},
}

@article{ahmad2022_ferrotrasmone,
  title = {Hybrid ferromagnetic transmon qubit: Circuit design, feasibility, and detection protocols for magnetic fluctuations},
  author = {Ahmad, Halima Giovanna and Brosco, Valentina and Miano, Alessandro and Di Palma, Luigi and Arzeo, Marco and Montemurro, Domenico and Lucignano, Procolo and Pepe, Giovanni Piero and Tafuri, Francesco and Fazio, Rosario and Massarotti, Davide},
  journal = {Phys. Rev. B},
  volume = {105},
  issue = {21},
  pages = {214522},
  numpages = {14},
  year = {2022},
  month = {Jun},
  publisher = {American Physical Society},
  doi = {10.1103/PhysRevB.105.214522},
  url = {https://link.aps.org/doi/10.1103/PhysRevB.105.214522}
}

@article{ahmad2024,
    author = {Ahmad, H. G. and Satariano, R. and Ferraiuolo, R. and Vettoliere, A. and Granata, C. and Montemurro, D. and Ausanio, G. and Parlato, L. and Pepe, G. P. and Tafuri, F. and Massarotti, D.},
    title = {Phase dynamics of tunnel {A}l-based ferromagnetic {J}osephson junctions},
    journal = {Applied Physics Letters},
    volume = {124},
    number = {23},
    pages = {232601},
    year = {2024},
    month = {06},
    issn = {0003-6951},
    doi = {10.1063/5.0211006},
    url = {https://doi.org/10.1063/5.0211006},
    eprint = {https://pubs.aip.org/aip/apl/article-pdf/doi/10.1063/5.0211006/19978492/232601_1_5.0211006.pdf},
}

@article{Connolly2024,
  title = {Coexistence of Nonequilibrium Density and Equilibrium Energy Distribution of Quasiparticles in a Superconducting Qubit},
  author = {Connolly, Thomas and Kurilovich, Pavel D. and Diamond, Spencer and Nho, Heekun and B\o{}ttcher, Charlotte G. L. and Glazman, Leonid I. and Fatemi, Valla and Devoret, Michel H.},
  journal = {Phys. Rev. Lett.},
  volume = {132},
  issue = {21},
  pages = {217001},
  numpages = {7},
  year = {2024},
  month = {May},
  publisher = {American Physical Society},
  doi = {10.1103/PhysRevLett.132.217001},
  url = {https://link.aps.org/doi/10.1103/PhysRevLett.132.217001}
}

@article{Aquino2025,
  title = {Nonequilibrium quasiparticles in superconducting circuits: Energy relaxation and charge and flux noise},
  author = {Nava Aquino, Jos\'e Alberto and de Sousa, Rog\'erio},
  journal = {Phys. Rev. Appl.},
  volume = {24},
  issue = {4},
  pages = {044088},
  numpages = {14},
  year = {2025},
  month = {Oct},
  publisher = {American Physical Society},
  doi = {10.1103/nr6y-zxnb},
  url = {https://link.aps.org/doi/10.1103/nr6y-zxnb}
}

@article{Wisne_2024,
doi = {10.1088/2633-4356/ad9a47},
url = {https://doi.org/10.1088/2633-4356/ad9a47},
year = {2024},
month = {dec},
publisher = {IOP Publishing},
volume = {4},
number = {4},
pages = {046001},
author = {Wisne, M and Deng, Y and Cansizoglu, H and Kopas, C and Mutus, J Y and Chandrasekhar, V},
title = {Transport signatures of phase fluctuations in superconducting qubits},
journal = {Materials for Quantum Technology}
}

@article{Iansiti,
  title = {Charging energy and phase delocalization in single very small {J}osephson tunnel junctions},
  author = {Iansiti, M. and Johnson, A. T. and Smith, Walter F. and Rogalla, H. and Lobb, C. J. and Tinkham, M.},
  journal = {Phys. Rev. Lett.},
  volume = {59},
  issue = {4},
  pages = {489--492},
  numpages = {0},
  year = {1987},
  month = {Jul},
  publisher = {American Physical Society},
  doi = {10.1103/PhysRevLett.59.489},
  url = {https://link.aps.org/doi/10.1103/PhysRevLett.59.489}
}

@article{PRL_Diagram,
  author = {Longobardi, Luigi and Massarotti, Davide and Stornaiuolo, Daniela and Galletti, Luca and Rotoli, Giacomo and Lombardi, Floriana and Tafuri, Francesco},
  title = {Direct Transition from Quantum Escape to a Phase Diffusion Regime in {Y}{B}a{C}u{O} Biepitaxial {J}osephson Junctions},
  journal = {Physical Review Letters},
  volume = {109},
  number = {5},
  pages = {050601},
  year = {2012},
  doi = {10.1103/PhysRevLett.109.050601},
}

@article{Stornaiuolo2013,
  title = {Resolving the effects of frequency-dependent damping and quantum phase diffusion in ${YBa}_{2}{Cu}_{3}{O}_{7-x}$ {J}osephson junctions},
  author = {Stornaiuolo, D. and Rotoli, G. and Massarotti, D. and Carillo, F. and Longobardi, L. and Beltram, F. and Tafuri, F.},
  journal = {Phys. Rev. B},
  volume = {87},
  issue = {13},
  pages = {134517},
  numpages = {7},
  year = {2013},
  month = {Apr},
  publisher = {American Physical Society},
  doi = {10.1103/PhysRevB.87.134517},
  url = {https://link.aps.org/doi/10.1103/PhysRevB.87.134517}
}

@article{evaluation_C,
author = {Maezawa,M.  and Aoyagi,M.  and Nakagawa,H.  and Kurosawa,I.  and Takada,S. },
title = {Specific capacitance of ${Nb/AlO_x/Nb}$ {J}osephson junctions with critical current densities in the range of 0.1–18 kA/cm$^{2}$},
journal = {Appl. Phys. Lett.},
volume = {66},
number = {16},
pages = {2134-2136},
year = {1995},
doi = {10.1063/1.113927},
}

@article{Krantz_2019,
   title={A quantum engineer’s guide to superconducting qubits},
   volume={6},
   ISSN={1931-9401},
   url={http://dx.doi.org/10.1063/1.5089550},
   DOI={10.1063/1.5089550},
   number={2},
   journal={Applied Physics Reviews},
   publisher={AIP Publishing},
   author={Krantz, P. and Kjaergaard, M. and Yan, F. and Orlando, T. P. and Gustavsson, S. and Oliver, W. D.},
   year={2019},
   month=jun }

@article{Cai_2022,
author = {Cai, Ranran and Žutić, Igor and Han, Wei},
title = {Superconductor/Ferromagnet Heterostructures: A Platform for Superconducting Spintronics and Quantum Computation},
journal = {Advanced Quantum Technologies},
volume = {6},
number = {1},
pages = {2200080},
doi = {https://doi.org/10.1002/qute.202200080},
url = {https://advanced.onlinelibrary.wiley.com/doi/abs/10.1002/qute.202200080},
eprint = {https://advanced.onlinelibrary.wiley.com/doi/pdf/10.1002/qute.202200080},
year = {2023}
}

@article{Senapati_2011,
  author    = {Kartik Senapati and Mark G. Blamire and Zoe H. Barber},
  title     = {Spin-filter {J}osephson junctions},
  journal   = {Nature Materials},
  year      = {2011},
  volume    = {10},
  pages     = {849--852},
  doi       = {10.1038/nmat3116},
  url       = {https://doi.org/10.1038/nmat3116}
}

@article{Confalone2025a,
  author    = {Tommaso Confalone and et al.},
  title     = {Preserving the {J}osephson Coupling of Twisted Cuprate Junctions via Tailored Silicon Nitride Circuits Boards},
  journal   = {Small},
  year      = {2025},
  volume    = {21},
  number    = {50},
  pages     = {2306220},
  doi       = {10.1002/smll.202306220}
}

@article{Confalone2025b,
  author    = {Tommaso Confalone and et al.},
  title     = {Cuprate Twistronics for Quantum Hardware},
  journal   = {Advanced Quantum Technologies},
  year      = {2025},
  volume    = {8},
  pages     = {2500203},
  doi       = {10.1002/qute.202500203}
}

@article{Marchegiani_2022,
  title = {Quasiparticles in Superconducting Qubits with Asymmetric Junctions},
  author = {Marchegiani, Giampiero and Amico, Luigi and Catelani, Gianluigi},
  journal = {PRX Quantum},
  volume = {3},
  issue = {4},
  pages = {040338},
  numpages = {22},
  year = {2022},
  month = {Dec},
  publisher = {American Physical Society},
  doi = {10.1103/PRXQuantum.3.040338},
  url = {https://link.aps.org/doi/10.1103/PRXQuantum.3.040338}
}

\end{document}